\documentclass[aps,prd,superscriptaddress,12pt,showpacs,notitlepage]{revtex4}

\usepackage{amsmath,amssymb,mathrsfs}
\usepackage[utf8]{inputenc}
\usepackage[dvips]{graphicx}
\usepackage{subfig}

\newcommand{\vecphi}{\vec \phi}
\newcommand{\vectau}{\vec \tau}
\newcommand{\trace}{\textrm{Tr}}
\newcommand{\total}{\textrm{d}}
\newcommand{\mathi}{\textrm{i}}
\newcommand{\mathe}{\textrm{e}}

\begin{document}
\title{Nonlinear field theory with topological solitons: Skyrme models}
\author{C. Adam}
\affiliation{Departamento de F\'isica de Part\'iculas, Universidad de Santiago de Compostela and Instituto Galego de F\'isica de Altas Enerxias (IGFAE) E-15782 Santiago de Compostela, Spain}
\author{C. Naya}
\affiliation{Departamento de F\'isica de Part\'iculas, Universidad de Santiago de Compostela and Instituto Galego de F\'isica de Altas Enerxias (IGFAE) E-15782 Santiago de Compostela, Spain}

\author{J. Sanchez-Guillen}
\affiliation{Departamento de F\'isica de Part\'iculas, Universidad de Santiago de Compostela and Instituto Galego de F\'isica de Altas Enerxias (IGFAE) E-15782 Santiago de Compostela, Spain}
\author{A. Wereszczynski}
\affiliation{Institute of Physics,  Jagiellonian University,
Reymonta 4, Krak\'{o}w, Poland}

\begin{abstract}

\noindent E-mail: adam@fpaxp1.usc.es, carlos.naya87@gmail.com, joaquin@fpaxp1.usc.es, wereszczynski@th.if.uj.edu.pl

\bigskip

In this talk, we give new insight into one of the best-known nonlinear field theories, the Skyrme model. We present some exact relevant solutions coming from different new versions (gauged BPS baby as well as vector BPS Skyrme models) giving rise to topological solitons, and highlighting the BPS character of the theory.
\end{abstract}

\maketitle

\section{Introduction}

In 1834, the Scottish naval engineer J. Scott Russell saw in the Edinburgh canal a solitary wave for the first time. Nowadays, they are known as solitons, and among their main features we can point out that they are localized solutions which conserve their form, even after collisions. One specific kind of these objects are the interesting topological solitons. They are given by the collective excitations of the relevant degrees of freedom of the system and their existence and stability are caused by the global topological structure of the base and field space.

It is here where the Skyrme model appears as a nonlinear field theory, introduced to describe nuclei, supporting topological solitons \cite{Skyrme1961, Skyrme1962, Skyrme1971}. The main feature of this model is that it considers the primary fields as pions whereas nucleons and nuclei are described by collective nonlinear excitations of the degrees of freedom, i.e., by topological solitons. One of the most clever ideas of Skyrme was to identify the topological charge which arises in the model with the baryon number ensuring the latter to be conserved. Although this is a theory in 3+1 dimensions, due to its nonlinear character and complexity, sometimes it is important to study its {\it little brother} in 2+1 too, the Baby Skyrme model \cite{Piette1995_1,Piette1995_2}, since it is simpler and can give us some clues to deal with the original model. This is an advice that we will follow in the present talk where two of the three different models are 2+1 dimensional theories.

The Standard Skyrme Model consists of two terms: a quadratic term in first derivatives (the nonlinear sigma-model term), and a quartic one (the so-called Skyrme term). Both of them are necessary for the stability of the solutions from the point of view of the Derrick scaling, however, we can generalize the model with two more terms asking to have terms at most quadratic in time derivatives (so we have a standard Hamiltonian formulation). Thus, we can add a potential and a sextic term, so the more general model is given by
\begin{equation}
\mathscr{L}_{Skr} = \mathscr{L}_2 + \mathscr{L}_4 + \mathscr{L}_6 + \mathscr{L}_0,
\end{equation}
with
\begin{equation} \nonumber
\mathscr{L}_2 = - \frac{f_\pi^2}{4} \trace(L_\mu L^\mu), \qquad \mathscr{L}_4=-\frac{1}{32 e^2} \trace \left([L_\mu,L_\nu]^2 \right),
\end{equation}
\begin{equation} \nonumber
\mathscr{L}_6 = \lambda^2 \pi^4 B_\mu^2, \qquad \mathscr{L}_0 = -\mu^2 V,
\end{equation}
and
\begin{equation}\nonumber
L_\mu = U^\dagger \partial_\mu U,
\end{equation}
\begin{equation}\nonumber
B^\mu = \frac{1}{24 \pi^2} \trace (\epsilon^{\mu \nu \rho \sigma} U^\dagger \partial_\nu U
 U^\dagger \partial_\rho U U^\dagger \partial_\sigma U),
\end{equation}
where $B^\mu$ is the topological current and the degrees of freedom are the $U$ fields ($U\in SU(2))$. This effective field theory is related to QCD in the large N (number of colours) limit, which does not favour any particular term. Another important comment on this Skyrme model is that although originally it was thought for Nuclear and Particle Physics, now it has applications in other branches, as for instance the planar Skyrmions in ferromagnetic materials: a collective excitation of spins from the homogeneous case.

Just to conclude this brief introduction we will present a key concept in our work, the BPS Bound, named after Bogomol'nyi, Prasad and Sommerfield. It consists in a lower energy bound of the system depending on its topology and not on the field configuration. Solutions saturating the bound (BPS solutions) fulfill a set of first order equations called BPS equations. Thus, BPS solutions imply a simplification of the field equations going from second to first order. Furthermore, when they saturate the bound, they minimize the energy, so BPS states are stable. An important characteristic for the Skyrme model is that the bound is related to topological quantities as the winding number: $E\propto n$. Unfortunately, the Standard Skyrme model cannot have solutions saturating the bound. One case where this is possible is the BPS Skyrme Model \cite{ASW} where we just take into account the sextic as well as the potential term:
\begin{equation}
\mathscr{L}_{BPS} = \mathscr{L}_6 + \mathscr{L}_0.
\end{equation}
This introduces the idea of infinite massive (quenched) pions since we forget about the kinetic term $\mathscr{L}_2$. This restricted Skyrme model solves old problems of hadrons and nuclei, and is being actively used in their phenomenology \cite{ASW2,BM,BM2}

With all these concepts in mind we are going to introduce three different new versions of the Skyrme model. In section 2, we will present a BPS baby model after introducing the gauge group U(1) with the usual Maxwell term for the gauge field. On the other hand, in section 3 we study the Skyrme field coupled to the vector meson $\omega$, firstly in 2+1 dimensions to finally reach the 3+1 dimensional case.


\section{Gauged BPS baby Skyrme model}

The first model we are going to deal with is the gauged version of the BPS Skyrme Model above \cite{ASW} but in 2+1 dimensions, the Gauged BPS baby Skyrme model \cite{paper1}. The Lagrangian density is given by
\begin{equation}
\mathscr{L} = -\frac{\lambda^2}{4} (D_\alpha \vecphi \times D_\beta \vecphi)^2
 - \mu^2 V(\vec{n} \cdot \vecphi) - \frac{1}{4 g^2} F^2_{\alpha \beta}.
\end{equation}
The degrees of freedom of the system are described by a vector of scalar fields $\vecphi=(\phi_1, \phi_2, \phi_3)$ with unit length $\vecphi^2=1$ and where the first term is quartic in first derivatives and at the same time is the square of the topological current, so it plays the r\^ole of the $\mathscr{L}_4$ and $\mathscr{L}_6$ in the 3+1 version. Moreover, the $\mathscr{L}_2$ term is not necessary for stability. On the other hand, since we have included the coupling to the gauge field, as well as the typical Maxwell term the usual derivatives have been replaced by the covariant ones:
\begin{equation}
D_\alpha \vecphi = \partial_\alpha \vecphi + A_\alpha \vec{n} \times \vecphi.
\end{equation}
Notice that if we ask for finite energy field configurations in the full baby model (the $\mathscr{L}_2$ term included), $\vecphi(\vec{x}, t)$ has to approach its vacuum value independently of the direction so the base space $\mathbb{R}^2$ can be compactified to a two-sphere $S^2$, and the field $\vecphi$ is a map $S^2 \rightarrow S^2$ characterized by an integer winding number or topological degree
\begin{equation}
\textrm{deg}[\vecphi] = \frac{1}{4 \pi} \int \total \, ^2x \vecphi \cdot \partial_1 \vecphi \times 
\partial_2 \vecphi = k, \qquad k \in \mathbb{Z}.
\end{equation}
Although this is not necessary for the BPS baby model we will assume it too because of consistency.

The next step, in order to simplify things, will be to choose $\vec{n}=(0, 0, 1)$ and the standard static ansatz:
\begin{equation}
\vecphi(r,\varphi) = \left( \begin{array}{c}
\sin f(r)\cos n \, \varphi \\
\sin f(r) \sin n \, \varphi \\
\cos f(r) \end{array} \right)
\end{equation}
\begin{equation}
A_0 = A_r = 0, \qquad A_\varphi= n a(r),
\end{equation}
so the electric and magnetic fields are
\begin{equation}
E_i = 0, \qquad B = \partial_1 A_2 - \partial_2 A_1 = \frac{n a'(r)}{r},
\end{equation}
whereas the winding number is just $\textrm{deg}[\vec \phi] = n$. It will be also useful to introduce the new variable and field
\begin{equation}
y=r^2/2, \qquad h=\frac{1}{2}(1-\cos f).
\end{equation}
In this model, we will work with the so called old potential: $V_o=1-\phi_3 = 2h$, a 2+1 version of the standard Skyrme potential used in 3+1 dimensions. Then, the boundary conditions are
\begin{equation}
h(0) = 1 \Leftrightarrow f(0) = \pi, \qquad a(0)=0
\end{equation}
at the origin, and if we have compacton solutions
\begin{equation}
a_y(y=y_0)=0, \qquad h(y_0)=h_y(y_0)=0
\end{equation}
at the radius, or
\begin{equation}
\lim_{y \to \infty} h(y) = 0, \qquad \lim_{y \to \infty} a_y(y) = 0
\end{equation}
for exponentially or powerlike localized solutions.  It can be shown that the old baby potential corresponds to compacton solutions. Although it seems we have five conditions notice that we are including a new parameter, the compacton radius $y_0$.

Before going to the solutions, we will ask ourselves if a BPS bound can be found for this model. Fortunately, the answer is positive. Thus, regarding the static energy functional and after a non-trivial manipulation we arrive at
\begin{equation}
E \geq E_0 \lambda^2 \int \total \, ^2 x q W' = 4 \pi |n| E_0 \lambda^2 \langle W' \rangle_{S^2} 
= 2 \pi |n| E_0 \lambda^2 |W(\phi_3=-1)|,
\end{equation}
where we will call $W$ the superpotential and $n$ the winding number (making clear it is a topological bound). This bound will be saturated if the fields obey the BPS equations given by
\begin{equation}
Q = W',
\end{equation}
\begin{equation}
B = -g^2 \lambda^2 W,
\end{equation}
where
\begin{equation}
Q \equiv \vecphi \cdot D_1 \vecphi \times D_2 \vecphi 
= q + \epsilon_{ij} A_i \partial_j (\vec{n} \cdot \vecphi) .
\end{equation} 
Let us comment on the superpotential $W$ present in this BPS bound. It appears in the process of deriving the bound that $W$ is not a free function but is restricted to globally existing (on the unit interval) solutions of the following superpotential equation
\begin{equation}
\lambda^2 W' \,^2 + g^2 \lambda^4 W^2 = 2 \mu^2 V.
\end{equation}
We give it this name since it is very similar to the superpotential equation of fake supergravity \cite{SUGRA}. In fact, we can find this kind of equation in self-gravitating domain walls, where the terms have different signs, or in extremal supersymmetric black holes, where they enter with the same sign. It is also necessary to remind that in order to have BPS solitons the solution of this superpotential equation has to exist globally. Since it depends on the potential $V$, we will not have a solution for all $V$. For instance, if V has two vacua there are no solitons. It is also important to remark that in this model all solutions of the static equations are BPS (we will see it with the numerical solutions).

\begin{figure}
 \begin{center}
  \subfloat[Function $h$ with its derivative (dashed line).]{\includegraphics[width=0.45\textwidth]{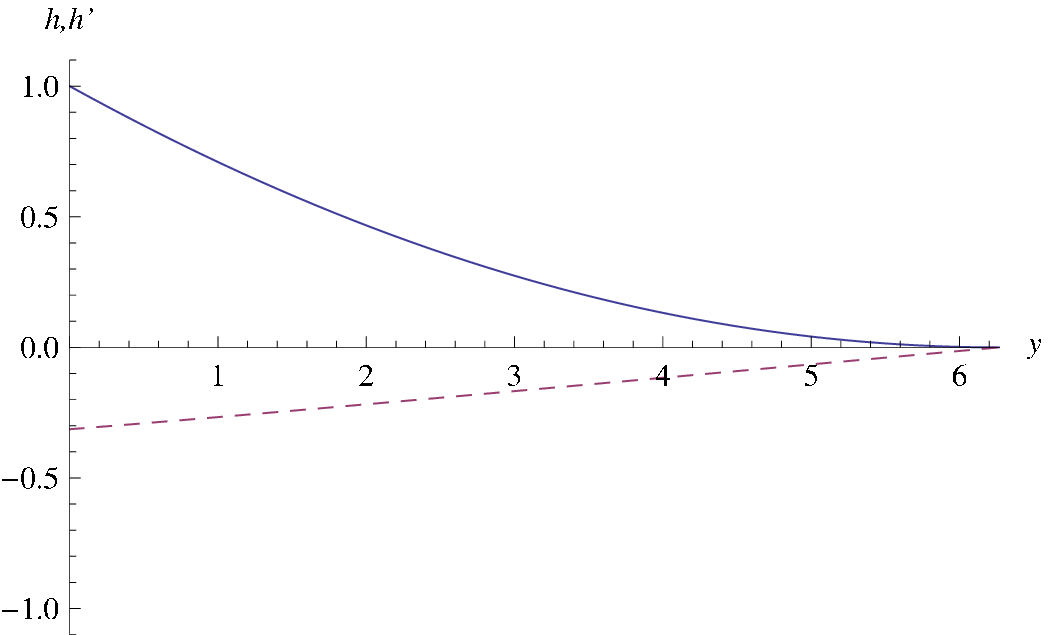}} \hspace{5mm} 
  \subfloat[Function $a$ and its derivative (dashed line).]{\includegraphics[width=0.45\textwidth]{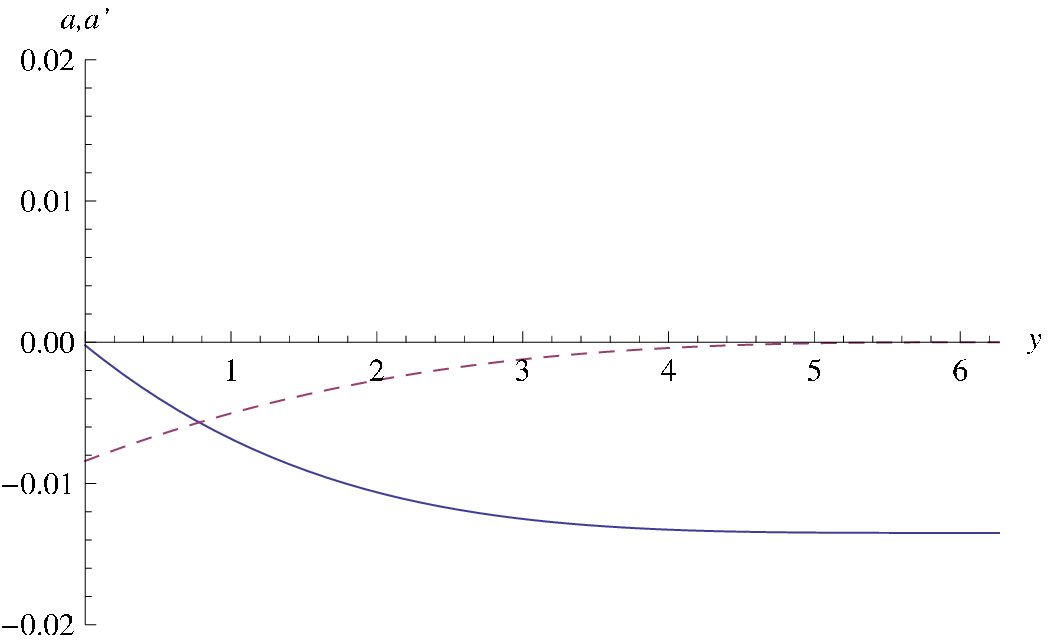}} 
  \caption{Solutions for $g=0.1$ and $\mu^2=0.1$.}
  \label{g01}
 \end{center}
\end{figure}
\begin{figure}
 \begin{center}
  \subfloat[Function $h$ with its derivative (dashed line).]{\includegraphics[width=0.45\textwidth]{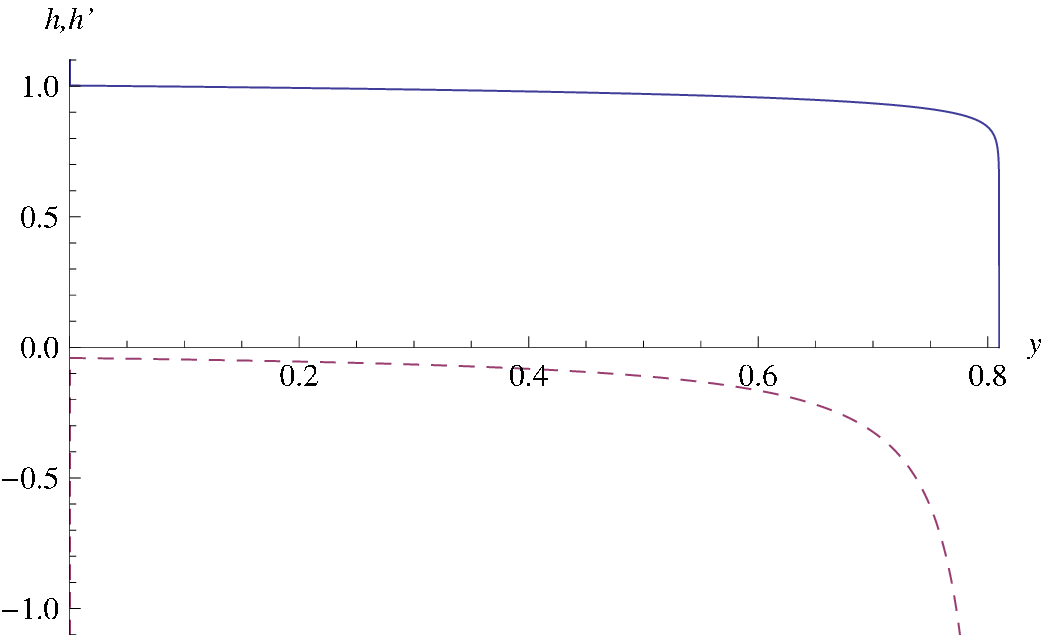}} \hspace{5mm} 
  \subfloat[Function $a$ and its derivative (dashed line).]{\includegraphics[width=0.45\textwidth]{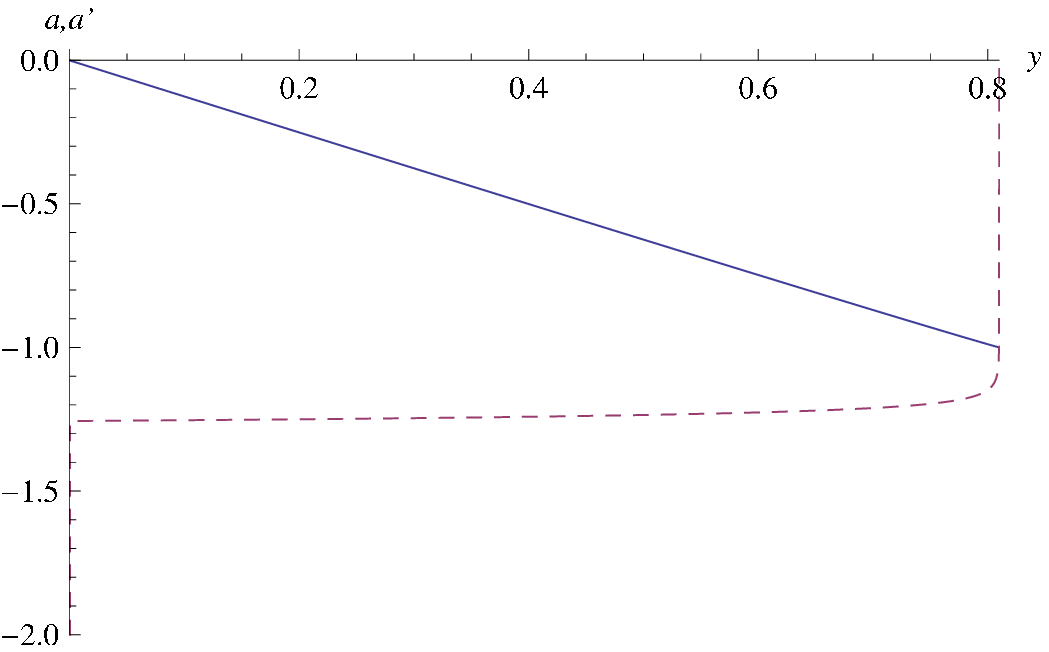}} 
  \caption{Solutions for $g=2$ and $\mu^2=0.1$.}
  \label{g2}
 \end{center}
\end{figure}
\begin{figure}
\begin{center}
\includegraphics[width=0.45\textwidth]{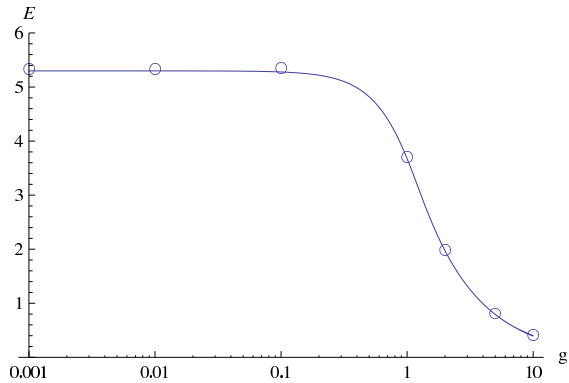}
\caption{Energy of static solutions (open circles) compared to the BPS bound (continuous line).}
\label{bound1}
\end{center}
\end{figure}

In figures \ref{g01} and \ref{g2} we can see the numerical solutions we have found for different values of the parameter $g$ (fixing $\lambda = n = 1$). In the case of the value of $\mu^2$ we just show cases corresponding to $\mu^2 = 0.1$ since the situation is quite similar for other values. However, this does not happen when considering $g$. In figure \ref{g01} we see the solution with $g=0.1$ (similar situation has been found for $g$ lower than this.), where the derivative of the field $a$ corresponding to the magnetic field is here localized at the center. On the other hand, looking now at figure \ref{g2} we see a different situation, with the magnetic field being almost constant until the compacton boundary and the field $h$ approaching the zero value suddenly. In fact, the higher the value of $g$, the more pronounced is the approach to zero.
Finally, using the numerical solutions for several values of $g$ we can see that indeed all soliton solutions are BPS as shown in figure \ref{bound1}, where the solid line represents the calculated bound $E_B=2\pi n \lambda^2 |W(h=1)|$.


\section{Vector Skyrme models}

In the model above we have just partially answered the question of how solitons interact with the lightest fields. We say partially because besides the U(1) Maxwell field we can consider another one to couple to solitons: the omega vector mesons, which after pions, are the lightest ones and their (phenomenologically motivated) coupling is just $\omega_\mu B^\mu$. As we have commented before, the 2+1 version of the system can be a good starting point to then move on to the 3+1 dimensional case, and this will be the process we are going to follow here. Finally, since we are going to compare this vector model with its BPS brother without additional mesons, we will have to be aware of possible changes in the typical mass hierarchy: the BPS model is based on pions with infinite mass while here we are regarding that heavier mesons have finite mass.


\subsection{Vector BPS baby Skyrme model}

As it was commented above, the difference of this model with respect to the first one is just that now we couple the solitons to a vector meson instead of a gauge field. Then, the 2+1 dimensional version of the Skyrme term is replaced here by the vector meson coupling \cite{paper2}:
\begin{equation} \label{lag_baby}
\mathscr{L} = - \mu^2 V(\phi_3) - \frac{1}{4}(\partial_\mu \omega_\nu- \partial_\nu \omega_\mu)^2
+ \frac{1}{2}M^2 \omega_\mu^2 + \lambda' \omega_\mu B^\mu,
\end{equation}
where $B^\mu$ is the topological current given by
\begin{equation}
B^\mu= - \frac{1}{8\pi} \epsilon^{\mu \alpha \beta} 
          \, \vecphi \cdot (\partial_\alpha \vecphi \times \partial_\beta \vecphi),
\end{equation}
and we will use the generalized old baby potentials
\begin{equation}
V= \left(\frac{1-\phi_3}{2} \right)^\alpha, \qquad \alpha \geq 1.
\end{equation}
As before, we can see the vector field $\vecphi$ as a map from $S^2$ to $S^2$ after compactification of the base space. Thus, we can use the stereographic projection
\begin{equation}
\vecphi = \frac{1}{1+|u|^2}(u+ \bar u, -i(u - \bar u), |u|^2-1),
\end{equation}
with the axially symmetric ansatz:
\begin{equation}
\quad u=f(r) \mathe^{\mathi n \phi} \qquad \omega \equiv  \omega=\omega(r) \qquad \omega_i = 0,
\end{equation}
where $n$ is the winding number.

The boundary conditions to have solitons will be
\begin{equation}\nonumber
f(r=0)= \infty, \qquad f(r=R)=0,
\end{equation}
\begin{equation}
\omega'(r=0)=0, \qquad \omega(r=R)=0,
\end{equation}
\begin{equation} \nonumber
f'(r=R)=0, \qquad \omega'(r=R)=0,
\end{equation}
where $R$ will be finite or infinite depending on the potential.

Studying now solutions for the model we will focus on the most interesting case, the potential with $\alpha = 2$. It is so interesting because we have an analytical solution as well as a BPS bound. Then, defining the new field $g=\frac{f^2}{1+f^2}$ and variable $x=r^2/2 = \frac{n \lambda}{\sqrt 2 \mu} y$ (where $\lambda$ is the initial $\lambda'$ up to a constant), we can solve the equation of motion for $g$
\begin{equation}
 g(y)= \frac{1}{K_1(2/\beta)} \frac{K_1(\frac{2}{\beta}\sqrt{1+\beta y})}{\sqrt{1+\beta y}} ,
\end{equation}
with $K_1$ the modified Bessel function of the second type and $\beta=\frac{2 \mu}{n \lambda M}$, whereas the equation for the vector meson is just
\begin{equation}
\omega_y = - \sqrt 2 \frac{\mu}{M}g.
\end{equation}
Finally, the BPS energy is
\begin{equation}
E = - \frac{\pi}{2} \Big( \frac{n \lambda}{\mu} \Big)^2 \omega(0) \omega_x(0) \simeq \sqrt 2 \pi \frac{n \lambda \mu}{M} \Big(1-\frac{1}{4} \beta +\cdots \Big),
\end{equation}
which is nonlinear on $n$. Therefore, states with high $n$ are unstable with respect to those with lower $n$. 
\begin{figure}
\begin{center}
\includegraphics[width=0.7\textwidth]{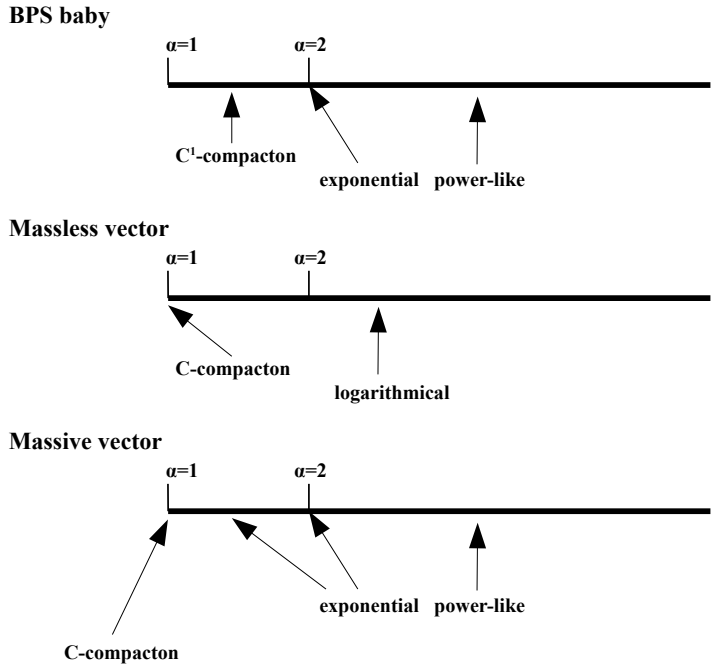}
\caption{Comparing solutions of the BPS baby Skyrme model with those of the baby vector one.}
\label{baby_sol}
\end{center}
\end{figure}

Other results either for massive vector mesons or massless ones are compared with those for the BPS baby model in figure \ref{baby_sol}. The most important difference appears for $\alpha=1$ since in this case, we get a C-compacton for the vector model. This means that at the boundary of the compacton there is a discontinuity, so that a Dirac delta is needed as a source term, what implies that the topological charge is screened. However, it is a positive difference since, because of the hierarchy problem, we have to remember that in the vector model the omega mesons are massless or have a finite mass whereas the lighter pions in nature are thought to be infinitely massive. Then, if the situation were the same the baby BPS Skyrme model would not be useful from the phenomenological point of view.

In the same way we could think about sending the mass of the omega mesons to infinity. With this we will recover the original mass hierarchy since if we assume that pions have infinite mass the same should stand for omegas. Then, our Lagrangian density (\ref{lag_baby}) would become
\begin{equation}
\mathscr{L} = -\mu^2 V(\phi_3) + \frac{1}{2}M^2 \omega_\mu^2 + \lambda' \omega_\mu B^\mu,
\end{equation}
so the field equation for $\omega$ mesons results in
\begin{equation}
\omega_\mu = -\frac{\lambda'}{M^2} B_\mu \quad.
\end{equation}
Plugging this back into the Lagrangian density we arrive at
\begin{equation}
\mathscr{L}= -\mu^2 V(\phi_3) -\frac{1}{2} \frac{\lambda'\,^2}{M^2} B_\mu^2.
\end{equation}
Thus, in this limit of infinite massive vector mesons we actually get the BPS baby Skyrme model.


\subsection{Vector BPS Skyrme model}

The next and natural step is to use the knowledge we have from the baby model for the 3+1 dimensional  version. Now, the corresponding Lagrangian density is \cite{paper3}:
\begin{equation}
\mathscr{L} = -\mu^2 V(U,U^\dagger) - \frac{1}{4}(\partial_\mu \omega_\nu-\partial_\nu \omega_\mu)^2 
+\frac{1}{2}M^2 \omega_\mu^2 + \lambda' \omega_\mu B^\mu,
\end{equation}
where $B^\mu$ is the topological current given by
\begin{equation}
B^\mu = \frac{1}{24 \pi^2}\trace (\epsilon^{\mu \nu \rho \sigma} U^\dagger \partial_\nu U
       U^\dagger \partial_\rho U U^\dagger \partial_\sigma U),
\end{equation}
and as before we will use the generalized Skyrme potentials, which in 3+1 dimensions are:
\begin{equation}
V=\left( \frac{1-\trace U}{2}\right)^\alpha = (1-\cos \, \xi)^\alpha.
\end{equation}
Furthermore, similar to the situation we had in 2+1 dimensions, the corresponding fundamental field $U \in$ SU(2) defines now a map from the three-sphere in the compactified base space to the field space SU(2), a map which is  characterized by a winding number. Then, a suitable parametrization for this $U$ field is
\begin{equation}
U = \mathe^{\mathi \xi \vec n \cdot \vectau} = \cos \, \xi + \mathi \sin \, \xi \, \vec n \cdot \vectau,
\end{equation}
with
\begin{equation}
\vec n = \frac{1}{1+|u|^2}(u+\bar u, -\mathi(u-\bar u),1-|u|^2).
\end{equation}
And as before, in order to simplify things, we will assume the static ansatz:
\begin{equation}
\omega_0 \equiv \omega = \omega(r), \qquad \omega_i = 0, \qquad \xi=\xi(r), 
\qquad u=\tan \frac{\theta}{2} \mathe^{\mathi n \phi}.
\end{equation}
Therefore, the boundary conditions for these new variables $\xi$ and $\omega$ will be
\begin{equation} \nonumber
\xi(r=0)=\pi, \qquad \omega(r=R_0)=0,
\end{equation}
\begin{equation}
\omega_r(r=0)=0, \qquad \omega(r=R_0)=0,
\end{equation}
where $R_0$ can be finite or infinite.
\begin{figure}
\begin{center}
\includegraphics[width=0.6\textwidth]{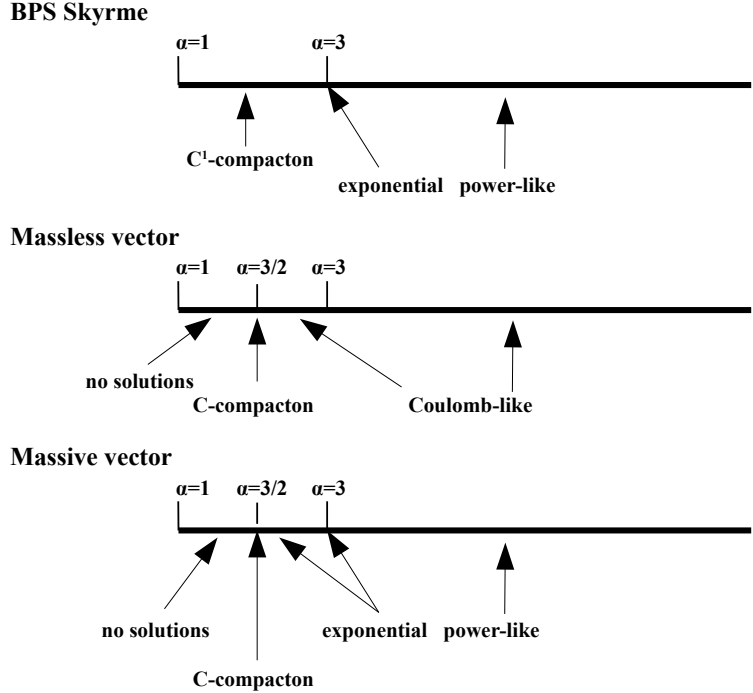}
\caption{Comparing solutions of the BPS Skyrme model with those of the vector one.}
\label{sol}
\end{center}
\end{figure}

Focusing now on the solutions of the massless and massive vector models, we can compare them with the BPS Skyrme model as we have done for the baby case (see figure \ref{sol}). Then, we can see that again C-compactons with the source term appear in the vector models for $\alpha=3/2$, but not for $\alpha=1$ where there are no solitons. This is an important issue since as we have commented above they are completely different models because of the mass hierarchy. However, in contraposition to what we found in the massive vector baby model, here we do not have a BPS solution for $\alpha = 2$, so why did we call this model BPS? The answer is that fortunately there exists a potential allowing BPS solitons both for massless and massive vectors, namely:
\begin{equation}
V_{BPS} = \frac{1}{4}(\xi-\cos \, \xi \sin \, \xi)^2,
\end{equation}
and the BPS energy calculated is
\begin{equation}
E =\frac{\pi^2 n^2 \lambda^2}{2} \omega(0),
\end{equation}
which for the massless case is analytically solvable getting
\begin{equation}
E=\frac{\pi^4}{8\sqrt{2 \sqrt 2}} \lambda \sqrt \frac{\lambda}{\mu} n^{3/2}.
\end{equation}
On the other hand, from analytical solutions for finite mass mesons we saw that in this case the energy behaves as $E \sim n$. So for both cases we have the same problem of unstable solitons for high $n^{6/5}$ we had in the baby model.


\section{Conclusions}

In this talk we have presented three new different BPS versions of the Skyrme model. In all of them we studied how solitons interact with the lightest fields: gauge and vector meson fields. Specially in the vector meson model we have made use of the simpler baby version of the model to help us with its generalization to the 3+1 case. In the case of the gauge field, this generalization is a more difficult task which lies outside the scope of this work.

Another important and interesting characteristic of all models is that BPS bounds have been discovered for them. In the case of the gauge model we have seen that a rich structure supports this topological bound with a superpotential equation similar to that in fake supergravity. In this case, BPS solitons exist for monotonically growing potentials and even more, all the solutions are BPS. Regarding the vector meson models, this BPS bound only exists for a specific potential: a generalized old baby potential in 2+1 dimensions and a similar to the generalized Skyrme potential for the 3+1 case.

Finally, it is necessary to emphasize that for some potentials the vector BPS models differ qualitatively from the BPS ones. It is a really important difference for the case of the standard pion mass potential, since in the BPS models we are assuming pions with infinite mass whereas in the vector ones we are adding omega mesons with finite mass, so we are changing the mass hierarchy. Therefore, if we had obtained different results, the BPS model would not be suitable from the phenomenological point of view.

\vspace*{0.3cm}

{\centerline {\bf Acknowledgement}}

\vspace*{0.2cm}

The authors acknowledge financial support from the Ministry of Education, Culture and Sports, Spain (grant FPA2008-01177), the Xunta de Galicia (grant INCITE09.296.035PR and Conselleria de Educacion), the Spanish Consolider-Ingenio 2010 Programme CPAN (CSD2007-00042), and FEDER. CN thanks the Spanish Ministery of Education, Culture and Sports for financial support (grant FPU AP2010-5772). Further, AW was supported by polish NCN grant 2011/01/B/ST2/00464.

\end{document}